\documentclass{aa}  

\usepackage{graphicx}
\usepackage{amsmath}
\usepackage{widetext}
\usepackage{cuted}
\usepackage[colorlinks=true, citecolor=blue, linkcolor=blue, urlcolor=blue, backref=page]{hyperref}
\usepackage{natbib}
\usepackage{txfonts}

\begin{document}

   \title{The effect of foreground galaxies on the estimation of the Hubble constant from type Ia supernovae}

   \subtitle{}

   \author{Amalia Villarrubia Aguilar
        \inst{1}\fnmsep\inst{2}\fnmsep\thanks{amalia.villarrubia-aguilar@apc.in2p3.fr}
          \and
          Julian Adamek
          \inst{2}\fnmsep\thanks{julian.adamek@uzh.ch}
          }

   \institute{Universit\'e Paris Cit\'e, CNRS, Astroparticule et Cosmologie, F-75013 Paris, France \and Institut f\"ur Astrophysik, Universit\"at Z\"urich, Winterthurerstrasse 190, 8057 Z\"urich, Switzerland\\}

   \date{Received February 2nd, 2025; accepted September 24, 2025}
   
  \abstract
   {Type Ia supernovae are the established standard candle in the construction of the Hubble diagram out to high-luminosity distances. The Hubble constant that fits observations of these supernovae best often turns out to be higher than fits to other data, which are therefore currently investigated for possible systematic effects. Many 
   studies focused on the calibration of supernova measurements either through the distance ladder in the local Universe or through large-scale anchors such as baryon acoustic oscillations or distant clusters. These different calibration methods yield different values for the Hubble constant, which means that calibration is a crucial step in the analysis of supernova measurements.}
   {We present a simulation-based assessment of another type of systematic effect that is related to the possibility that the line of sight to a distant supernova passes close to a foreground galaxy.}
   {We considered two cases separately: First, the foreground galaxy may block the line of sight so that the supernova is not observed. Since foreground galaxies are correlated with overdensities that typically magnify the flux of background sources, this effect systematically removes lensed supernovae from the sample and biases the high-redshift Hubble diagram towards demagnified (fainter) supernovae. Second, if the supernova can be observed, its proximity to the foreground galaxy can lead to an incorrect host assignment, especially when the true host has a low surface brightness. Since foreground galaxies are typically found at lower redshifts, this effect introduces another systematic bias. The probability of line-of-sight alignments with foreground galaxies increases with redshift and therefore affects distant supernovae more strongly.}
   {We find that these two effects are both weak, but the effect of host misidentification should be included in the systematic error budget at the current levels of measurement precision.}
   {}


\titlerunning{The effect of foreground galaxies on the estimation of the Hubble constant from type Ia supernovae}
\authorrunning{A. Villarrubia Aguilar \& J. Adamek}

   \maketitle

\section{Introduction}
Since Hubble’s initial measurements, our methods for estimating the Hubble constant ($H_0$) have developed significantly and incorporate more precise observational and data analysis techniques as well as refined theoretical models. These improvements have led to the emergence of the so-called Hubble tension, which is a persistent statistically significant discrepancy between the Hubble constant values obtained through different estimation methods.

In the late Universe, the leading method for measuring $H_0$ is based on the brightness-redshift correlation of type Ia supernovae (SNe Ia). The latest measurement from the \textit{Pantheon}$+$ analysis, carried out by the \textit{SH0ES}\footnote{\textit{supernovae and $H_0$ for the equation of state of dark energy }} team, yielded a value of $H_0 = 73.5 \pm 1.1\,\mathrm{km}\,\mathrm{s}^{-1}\,\mathrm{Mpc}^{-1}$ \citep{Brout2022}. On the other hand, the gold standard of the early Universe constraint was obtained by the \textit{Planck} team in 2018, who reported a value of $H_0 = 67.27 \pm 0.60\,\mathrm{km}\,\mathrm{s}^{-1}\,\mathrm{Mpc}^{-1}$ at the $68\%$ confidence level, assuming a six-parameter $\Lambda$ cold dark matter ($\Lambda$CDM) model \citep{Planck2020}. The discordance between different $H_0$ measurements, which ranges from $4\sigma$ to $6\sigma$ depending on the specific dataset used, is considered to be one of the most long-lasting and widely persistent challenges faced by contemporary cosmology \citep{DiValentino2021_review}. Comparing the value of $H_0$ measured locally to the value predicted from the early Universe measurements is an essential end-to-end test of the validity of the currently accepted $\Lambda$CDM model over the longest possible time span: from the dense dark-matter dominated early Universe to the current diluted dark-energy dominated Universe. In this context, the Hubble tension might be interpreted as a harbinger of new physics that indicates the partial failure of the $\Lambda$CDM model. We lack a compelling alternative model, however, and it therefore remains relevant to search for new sources of biases or systematic errors in our measurements of $H_0$.

We analyse the biases introduced in our late-time estimation of $H_0$ by the obstruction of certain SNe Ia by foreground galaxies that lie on the same line of sight in our observations. We study two different scenarios and assess how they each might affect our estimate of $H_0$. In the first scenario of total blocking, we assume that obstructed supernovae (SNe) are outshone by the foreground galaxies and are not visible to us. This leads to a selection bias in our observed SNe Ia catalogues. In the second scenario, we assume that the light from obstructed SNe is visible to us through their foreground galaxies. This leads to a different bias on our late-time $H_0$ estimation that is related to the method that is used to match each observed supernova (SN) to its host galaxy. This estimation relies on our ability to perform this matching correctly \citep{Carr2022}. This non-trivial task is currently being performed in a way that is not impervious to errors for obstructed SNe, however. To isolate these two effects from other observational systematics, we study them using a synthetic galaxy catalogue from a large N-body simulation that includes full ray tracing for each line of sight to accurately model weak gravitational lensing and redshift ($z$) perturbations.

The remainder of this paper is structured as follows. Section \ref{sec:dataset} presents the catalogue of simulated galaxies and SNe Ia we used for this analysis. The method for estimating $H_0$ from a simulated SNe Ia sample is discussed in Section \ref{sec:method}. In Section \ref{sec:results} we present our results. In particular, Section \ref{sec:total_blocking} is dedicated to discussing the impact of the selection bias from total blocking on our $H_0$ estimation, while we assess the impact of host galaxy misidentification for obstructed SNe Ia on our $H_0$ estimation in Section \ref{sec:mismatching}. Finally, we conclude in Section \ref{sec:conclusion}.

\section{\label{sec:dataset}The synthetic dataset}
In this section, we describe how we constructed a catalogue of simulated SNe. We then discuss the statistics of supernova blocking and host galaxy mismatching derived from this catalogue.

\subsection{\label{sec:supernova_catalog} The synthetic supernova catalogue}
This analysis is based on a dark matter halo catalogue extracted from a relativistic N-body simulation performed with the code \texttt{gevolution} \citep{Adamek2016_gevolution}. The simulation was run assuming a flat $\Lambda$CDM cosmology with $H_0 = 67\,\mathrm{km}\,\mathrm{s}^{-1}\,\mathrm{Mpc}^{-1}$, $\Omega_\mathrm{m} = 0.319$ (which includes two massive neutrino species in a minimum-mass scenario, $\sum m_\nu \approx 0.06\,\mathrm{eV}$) and a radiation density that accounted for one effectively massless neutrino species. The linear initial conditions were generated using \texttt{CLASS}\footnote{\textit{{cosmic linear anisotropy solving system}}} \citep{Blas2011} at an initial redshift of $z_{\text{ini}} = 127$, based on a primordial power spectrum with amplitude $A_\mathrm{s} = 2.215 \times 10^{-9}$ (evaluated at the pivot scale $0.05\, \mathrm{Mpc}^{-1}$) and spectral index $n_\mathrm{s} = 0.9619$. The simulation used a periodic box with side length $1440\,h^{-1}\,\mathrm{Mpc}$, containing $5760^3$ particles. The gravitational potential was sampled at a spatial resolution of $250\,h^{-1}\,\mathrm{kpc}$.

The halo catalogue, constructed from the simulated light cone with the \texttt{ROCKSTAR}\footnote{\textit{robust overdensity calculation using K-space topologically
adaptive refinement}} halo finder \citep{Behroozi2012}, included about 17 million haloes in the redshift range $0.001 < z < 1.27$. At low redshift ($z \leq 0.1$, corresponding to a comoving distance of $292\,h^{-1}\,\mathrm{Mpc}$), the haloes spanned the full sky volume. Beyond this redshift threshold, the simulated objects spanned a cone-shaped volume covering an area of $1000$ square degrees. The observed redshift and luminosity distance of each halo were determined by integrating the geodesic equation and Sachs equations along its light ray, using the gravitational potential along the line of sight, and accounting for the Doppler contribution due to the peculiar motion of the source \citep{Adamek2019}. This method does not assume any fixed distance-redshift relation, but calculates the true observed values for each source separately. The catalogue also included a mass proxy that was based on the number of N-body particles assigned to each halo.

To sidestep the complications of simulating baryonic matter, we assumed that there is a one-to-one correspondence between dark matter haloes and galaxies: Each of the dark matter haloes in the catalogue contained a single galaxy in which a SN Ia could be observed. To assign a physical radius to each of these galaxies, we used an abundance-matching approach that relates the mass distribution of dark matter haloes retrieved from the simulation to the radius distribution of galaxies observed in large galaxy surveys. We used an observed catalogue of $50\,000$ galaxies from the \textit{Sloan Digital Sky Survey} \citep[SDSS,][]{SDSS7} with measured half-light radii. The catalogue was drawn from a $30^\circ \times 30^\circ$ field around the north Galactic pole, excluding low-quality sources for which the radius error exceeded one-third of the measured radius. To ensure completeness, we additionally only selected galaxies with redshift $z < 0.4$ and assumed that the radius distribution was independent of redshift. It is known that galaxy sizes tend to grow over time \citep{vanderWel:2014wba}, and neglecting this effect is therefore a limitation of our modelling. In the context of the selection biases we are interested in, however, this is a conservative choice in the sense that accounting for this effect would tend to decrease the biases we study further.

Our abundance matching relied on the assumption that the total number of galaxies and dark matter haloes per unit volume are equal (each halo hosts one galaxy) and that there is a monotonic relation between halo mass and galaxy radius. Although the simulated and observed catalogues that we sought to relate have different unrelated lower thresholds, the most massive haloes and largest galaxies were expected to be included in both catalogues. Therefore, the galaxies with the largest radii can be associated with the haloes with the highest masses by equating their cumulative number densities,
\begin{equation}
    \int \limits_{r(m)}^{\infty} n_{\text{R}}(r') \, dr' = \int \limits_{m}^{\infty} n_{\text{M}}(m') \, dm'\,, 
\end{equation}
where $n_{\text{R}}(r)$ is the number density of galaxies per radius interval, and $n_\text{M}(m)$ is the number density of haloes per mass interval. This relation implicitly defines the galaxy radius as a function of the halo mass $r(m)$, allowing us to assign a radius to the galaxy that is contained in each simulated dark matter halo. To also assign a S\'ersic index to each simulated galaxy, we used the relation between the half-light radius $r_{\text{HL}}$ and S{\'e}rsic index $n$ derived by \citet{Graham2013},
\begin{multline}
    \log_{10} (r_{\text{HL}} [\mathrm{kpc}]) = \ 0.434 n - 0.94 \log_{10} (n)\\
+  \frac{0.5 b^{2n}}{e^b n \Gamma (2n)} - 0.364 \,,
\end{multline}
for $0.5 < n < 10$ with $ b \approx 1.9992n - 0.3271$ \citep{Capaccioli1989}.\footnote{The expression for $b$ can be derived from Eq.~\eqref{eq:sersic} by requiring that half of the light comes from $r < r_\mathrm{HL}$.} This relation can be inverted for the radii of almost all galaxies in our catalogue and typically yielded a S\'ersic index between 3 and 6. For the small number of galaxies for which $n$ had no solution, we fixed it to the value $n=0.5$. This occurred in fewer than $0.055\%$ of the cases.

Finally, the algorithm we used to place a SN into each of these galaxies consisted of two main steps: First, we chose the separation $r_{\text{sep}}$ (in parsecs) between the SN and the centre of its host galaxy, and then we determined the SN coordinates accordingly. For the first step, we relied on the results from \citet{Kelly2008}, who indicated that the distribution of SNe Ia within their host galaxies follows the respective intensity profile. Galaxy intensity profiles are typically described by a S\'ersic profile \citep{Sersic1963}, which gives the intensity, $I$, as a function of the distance from the galactic centre, $r$,
\begin{equation} 
\label{eq:sersic}
I(r) \propto \exp \left[ -b \left( \frac{r}{r_{\text{HL}}} \right)^{1/n} \right]\,.
\end{equation}
We used the S\'ersic profile of each galaxy as the probability distribution function from which we drew $r_{\text{sep}}$, that is, $p(r_\mathrm{sep}) dr_\mathrm{sep} \propto r I(r) dr$, which was then converted into an angular separation $\alpha_{\text{sep}}$. Using $\alpha_{\text{sep}}$ and the coordinates of the galaxy centre $(\theta_{\text{G}}, \phi_{\text{G}})$, we computed provisional SN coordinates $(\theta_{\text{SN}}, \phi_{\text{SN}})$, initially assuming $\theta_{\text{SN}} = \theta_{\text{G}}$. The final SN coordinates were obtained by rotating $(\theta_{\text{SN}}, \phi_{\text{SN}})$ around the vector pointing to the galaxy centre by a random angle $\beta$ ($0^{\circ} \leq \beta < 360^{\circ}$) to ensure that the SN was randomly oriented relative to its host galaxy. These steps result in a synthetic catalogue of SNe Ia and their corresponding host galaxies, which we used for the analysis of line-of-sight obstruction and host galaxy mismatching presented in the remainder of this paper.

\subsection{\label{sec:blocking_statistics} Total blocking scenario}
In its simplest form, total blocking implies that SNe lying on the same line of sight as a galaxy in their foreground were excluded from our observed catalogues. This first scenario assumes that the light from these SNe is sufficiently absorbed by the foreground galaxy to prevent their detection. We determined which SNe in our simulated catalogue are affected by this blocking based on two straightforward criteria. First, the blocking galaxy had to lie in the foreground of the SN. Second, a SN was blocked by a galaxy lying in its foreground when the angular separation between the SN and the galaxy, $\theta$, was smaller than the angular half-light radius of the galaxy, $\alpha_{\text{HL}}$.
This means
\begin{eqnarray}
    d_{C}^{\text{G}} &<& d_{C}^{\text{SN}}\,, \\
    \theta &<& \alpha_{\text{HL}}\,,
\end{eqnarray}
where $d_{C}^{\text{G}}$ and $d_{C}^{\text{SN}}$ correspond to the comoving distance of the galaxy and the SN, respectively. When multiple galaxies fulfilled these two conditions for the same SN, we considered that it was blocked by the galaxy for which $\theta$ is lowest.
\begin{figure}
    \centering
    \includegraphics[width=1\linewidth]{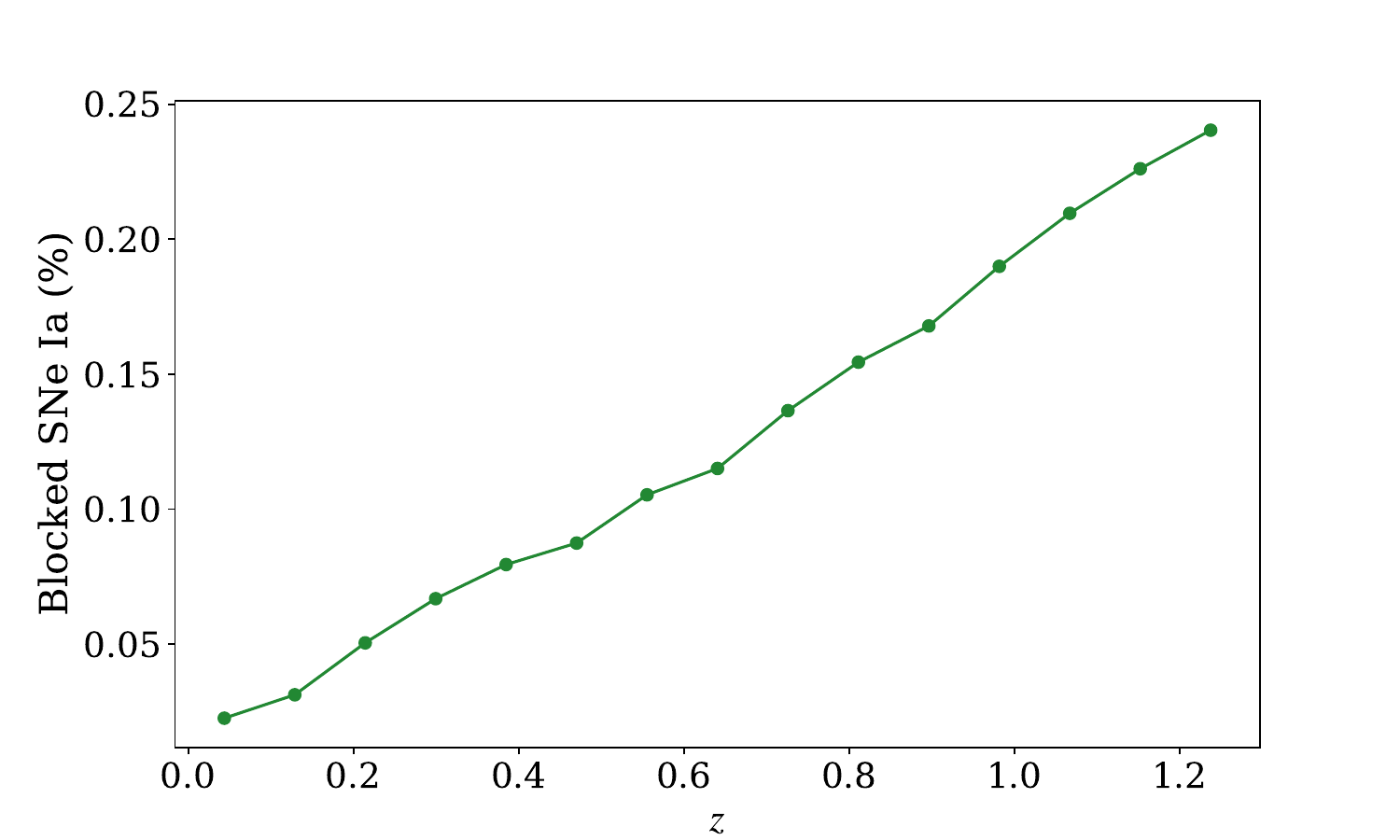}
    \caption{Percentage of obstructed SNe Ia in our synthetic catalogue as a function of redshift.}
    \label{fig:blocking_vs_z}
\end{figure}
We found that $0.16 \%$ of the SNe in our whole catalogue were blocked by a foreground galaxy according to the blocking criteria given above. The evolution of this percentage as a function of redshift is shown in Figure \ref{fig:blocking_vs_z}. As expected, the fraction of blocked SNe increases somewhat with redshift as the number of intervening galaxies grows, such that $0.22 \%$ of the SNe in our catalogue above $z = 1$ were blocked by a foreground galaxy.

\begin{figure}
    \centering
    \includegraphics[width=1\linewidth]{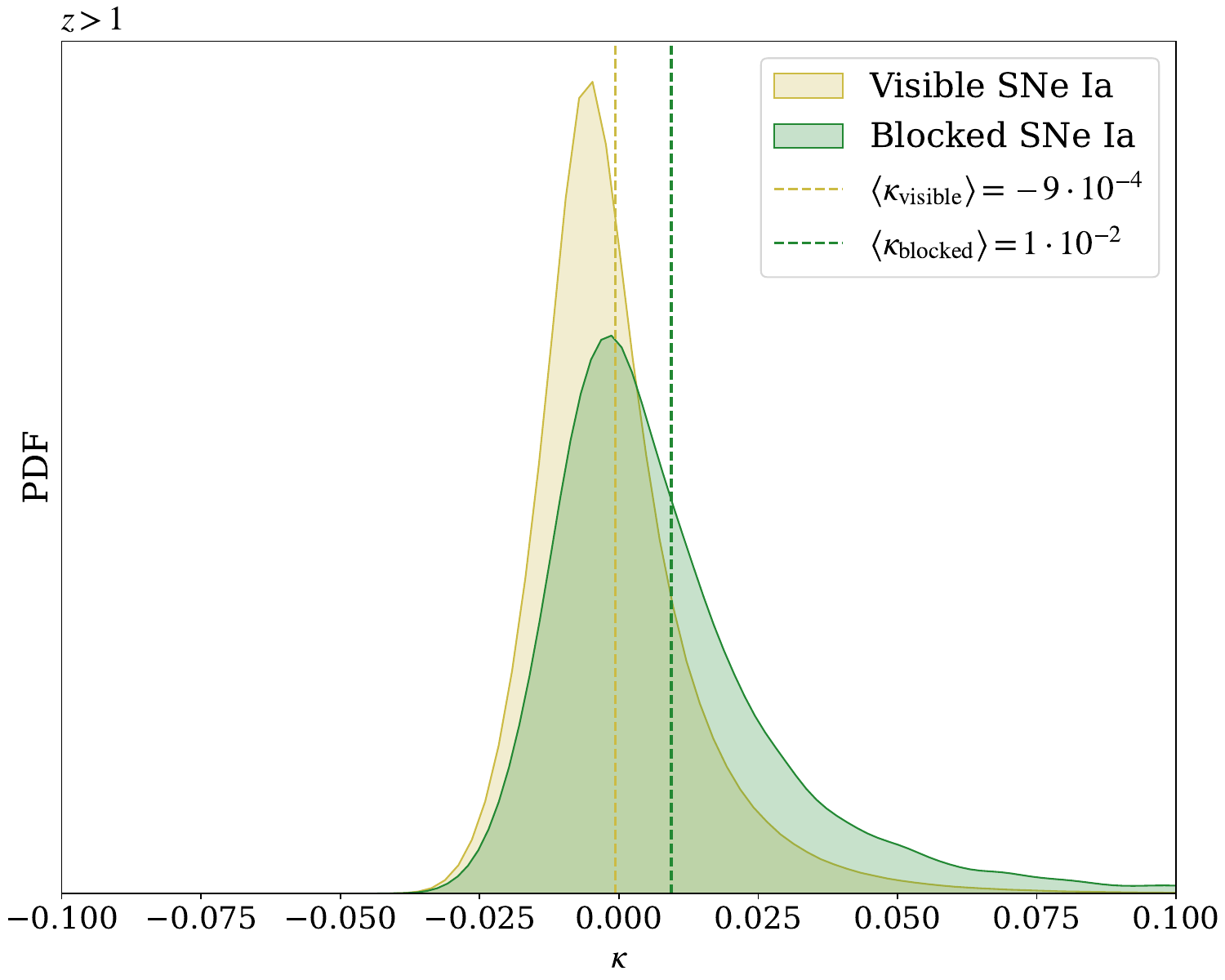}
    \caption{Distribution of the convergence parameter $\kappa$ for the blocked and the visible SNe Ia in our catalogue above redshift $z=1$.}
    \label{fig:blocking_kappa_dist}
\end{figure}

We investigated the effect of gravitational lensing on these blocked SNe using the convergence parameter $\kappa$ defined by \citep[for example in][]{Lepori2020} as 
\begin{equation}
    \kappa \equiv 1- \frac{d_A}{\bar{d_A}}\,,
\end{equation}
where $\bar{d}_A$ is the angular diameter distance to a source under the assumption of a Friedmann–Lemaître–Robertson–Walker (FLRW) metric, and $d_A$ is the measured angular diameter distance that we computed by ray tracing along each specific line of sight. The convergence parameter describes whether a source is magnified ($\kappa >0$) or demagnified ($\kappa <0$) by gravitational lensing. The average convergence for blocked SNe above $z=1$ is $\langle \kappa_{\text{blocked}} \rangle = 1 \times 10^{-2}$, compared to $\langle \kappa_{\text{visible}} \rangle = -9 \times 10^{-4}$ for visible SNe. As shown in Figure \ref{fig:blocking_kappa_dist}, the distribution of $\kappa$ is more skewed towards positive values for blocked sources, indicating that they are more magnified than their visible counterparts on average.

\begin{figure}
    \centering
    \includegraphics[width=\linewidth]{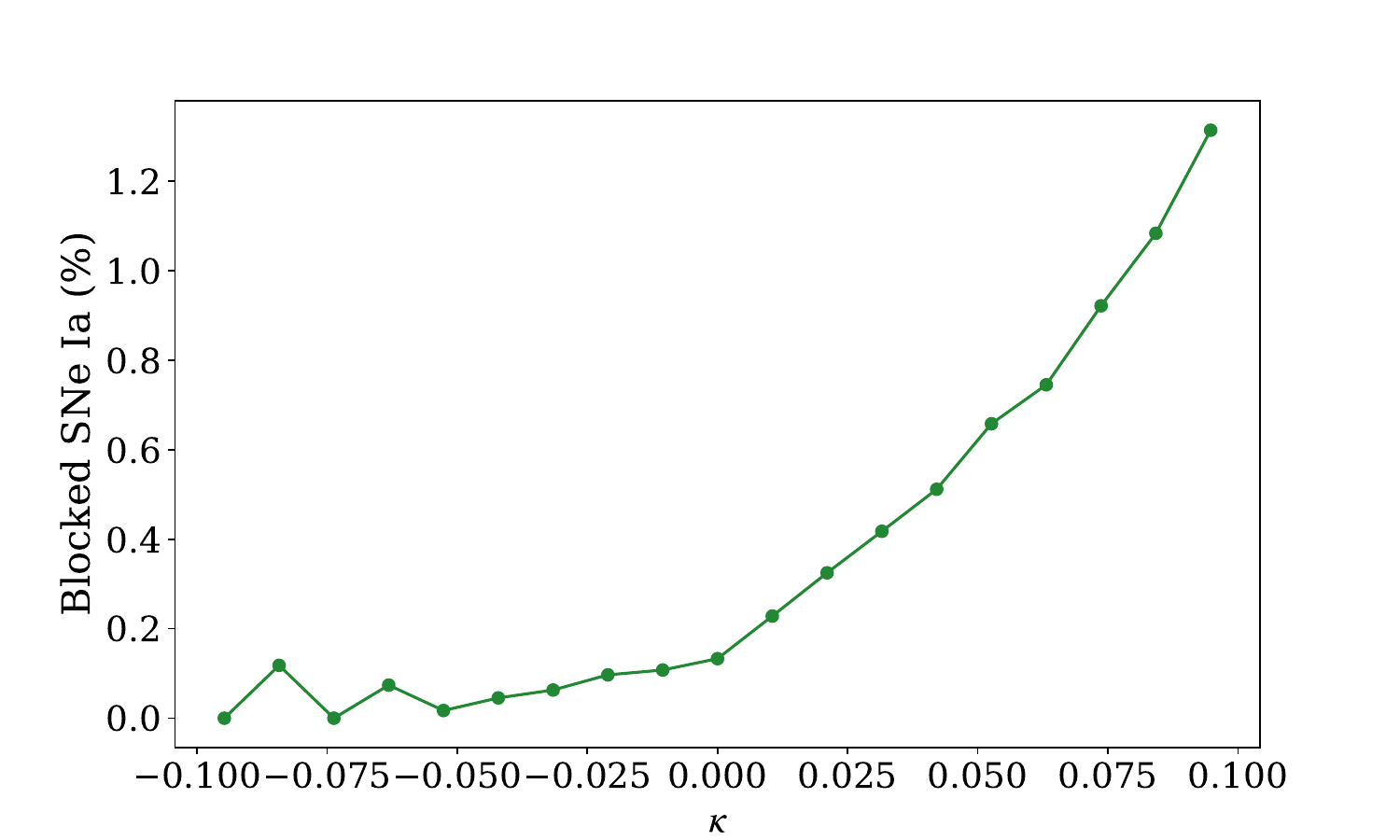}
    \caption{Percentage of obstructed SNe Ia in our synthetic catalogue as a function of the convergence parameter $\kappa$}
    \label{fig:blocking_vs_kappa}
\end{figure}

To illustrate this point further, Figure \ref{fig:blocking_vs_kappa} presents the percentage of SNe Ia that were blocked in our synthetic catalogue as a function of $\kappa$. The fraction of blocked SNe increases with $\kappa$, indicating that strongly magnified sources have a significantly higher probability of being obstructed. This trend is explained by the fact that SNe blocked by a foreground galaxy tend to lie along over-dense lines of sight. As a consequence of magnification, their distances tend to be underestimated ($d_L^{\text{obs}} < \bar{d}_L$). Additionally, because the fraction of blocked SNe Ia varies with redshift, these sources are not randomly distributed in the Hubble diagram. We assess whether this selection bias introduced by SN blocking affects our estimation of $H_0$ in Section \ref{sec:total_blocking}.

\subsection{\label{sec:mismatching_statistics} Incorrect host assignments}
The fact that certain SNe lie behind a foreground galaxy does not necessarily imply that they are all missing from our observations. At their peak, SNe Ia can reach luminosities comparable to those of an entire galaxy. This motivates the consideration of an alternative scenario in which these SNe remain visible despite the presence of a foreground galaxy along the line of sight. Unlike the total blocking scenario described in the previous section, here we considered the case when the foreground galaxies do not fully absorb the SNe light, allowing them to remain detectable.
This scenario has implications for the way in which we assign redshifts to the SNe. The two blocking scenarios outlined here and in Section \ref{sec:blocking_statistics} are therefore based on fundamentally different assumptions and must be considered separately. 

The preferred method is to determine SN redshifts indirectly using the spectroscopic redshift of their host galaxies, which have sharper spectral lines that allow for a higher accuracy. Correctly associating each SN with its corresponding host galaxy is therefore crucial \citep{Carr2022}. The algorithm we used to perform this matching for the SNe Ia catalogue used for the latest measurement of $H_0$ by the \textit{SH0ES} team was described by \citet{Gupta2016}. When no spectroscopic information is available for a SN, this matching algorithm consists of computing the ratio $d_{\mathrm{DLR}}$ for each galaxy in a radius of $30''$ around the SN. For a supernova-galaxy pair, the quantity $d_{\mathrm{DLR}}$ is defined as 
\begin{equation}
    d_{\text{DLR}} = \frac{\theta}{\alpha_{\text{DLR}}}\,,
\end{equation}
where $\theta$ corresponds to the angular separation between the SN and the centre of the galaxy, and the directional light radius $\alpha_{\text{DLR}}$ represents the elliptical radius of the galaxy in the SN direction. The galaxy for which $d_{\text{DLR}}$ is lowest is then identified as the host galaxy. This method is prone to errors for obstructed SNe, which have small angular separations from their obstructing galaxies. There is a risk of incorrectly associating these SNe with their obstructing galaxy rather than their true host galaxy, and thus, assigning them an incorrect redshift.
\begin{figure}
    \centering
    \includegraphics[width=1\linewidth]{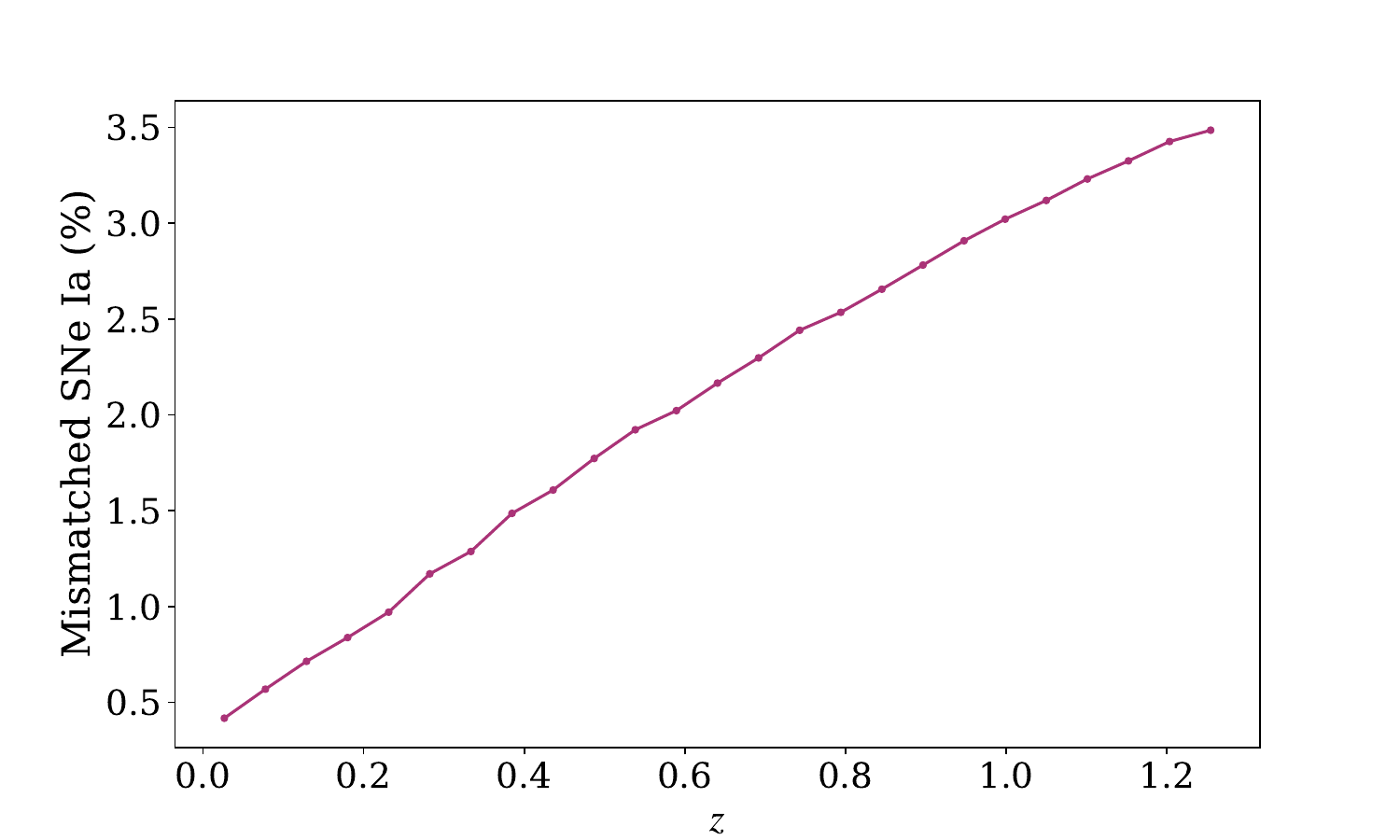}
    \caption{Percentage of SNe Ia that are mismatched to a foreground galaxy in our catalogue as a function of redshift.}
    \label{fig:mismatching_vs_z}
\end{figure}

Ideally, if all SNe Ia were observed spectroscopically, this would provide direct redshift measurements. These measurements might then be cross-checked against the redshift of their assumed host galaxies, which would help us to mitigate errors from host galaxy misidentification. Since the advent of large SN surveys such as the \textit{Supernova Legacy Survey} (SNLS) and the \textit{Dark Energy Survey} \citep[DES,][]{DES:2024jxu}, which contain thousands of supernovae, obtaining spectroscopic confirmation for each SNe Ia has become increasingly unfeasible, however. We therefore primarily relied on redshifts that were obtained indirectly from host galaxy spectra, and for the majority of supernovae, there is no direct redshift information to support a host identification \citep{Gupta2016}. For this reason, we assumed that our simulated supernovae were only observed photometrically and that the spectroscopic redshift of their host galaxy was obtained separately.

We applied the supernova - host galaxy matching algorithm described above to our simulated catalogue and found that $ 2.6 \%$ of the SNe in our whole catalogue were matched to a foreground galaxy that obstructed them rather than to their host galaxy. Figure \ref{fig:mismatching_vs_z}  shows how this percentage evolves with redshift: Again, the fraction of mismatched sources increases with redshift. Unlike \citet{Gupta2016}, we did not limit the host galaxy search to a $30''$ radius around each galaxy in our implementation. We only computed $d_{\text{DLR}}$ for the galaxies lying in the foreground of each SN, however, because we focuse on the effect of foreground galaxies. This restriction is somewhat arbitrary because similar host misidentifications might also occur with background galaxies if they fall within the thresholds of the survey in terms of flux and surface brightness. Moreover, we worked under the simplifying assumption that our simulated galaxies are circular, so that we were able to replace the directional light radius by the half-light radius in the matching criterion.

To evaluate the potential contribution of background galaxies, we also ran the matching algorithm without restricting it to foreground galaxies. Instead, we considered all galaxies with $d_L^{\mathrm{G}} < 1.25 \times d_L^{\mathrm{SN}}$ as possible host candidates for each SN. This luminosity distance threshold ensured that nearly all foreground galaxies, as well as a significant fraction of background galaxies, were included in the matching process. Although somewhat arbitrary, this luminosity distance cutoff accounts for the fact that galaxies at much larger luminosity distances than the SN would likely be undetected or excluded because they are inconsistent with the SN distance. In this scenario, the fraction of incorrect host galaxy assignments in our full catalogue increased to $3.2\%$. Therefore, the results for the host galaxy misidentification presented in the rest of this paper provide only qualitative insights, and a precise quantification of the bias requires a more realistic treatment.

\begin{figure}
    \centering
    \includegraphics[width=1\linewidth]{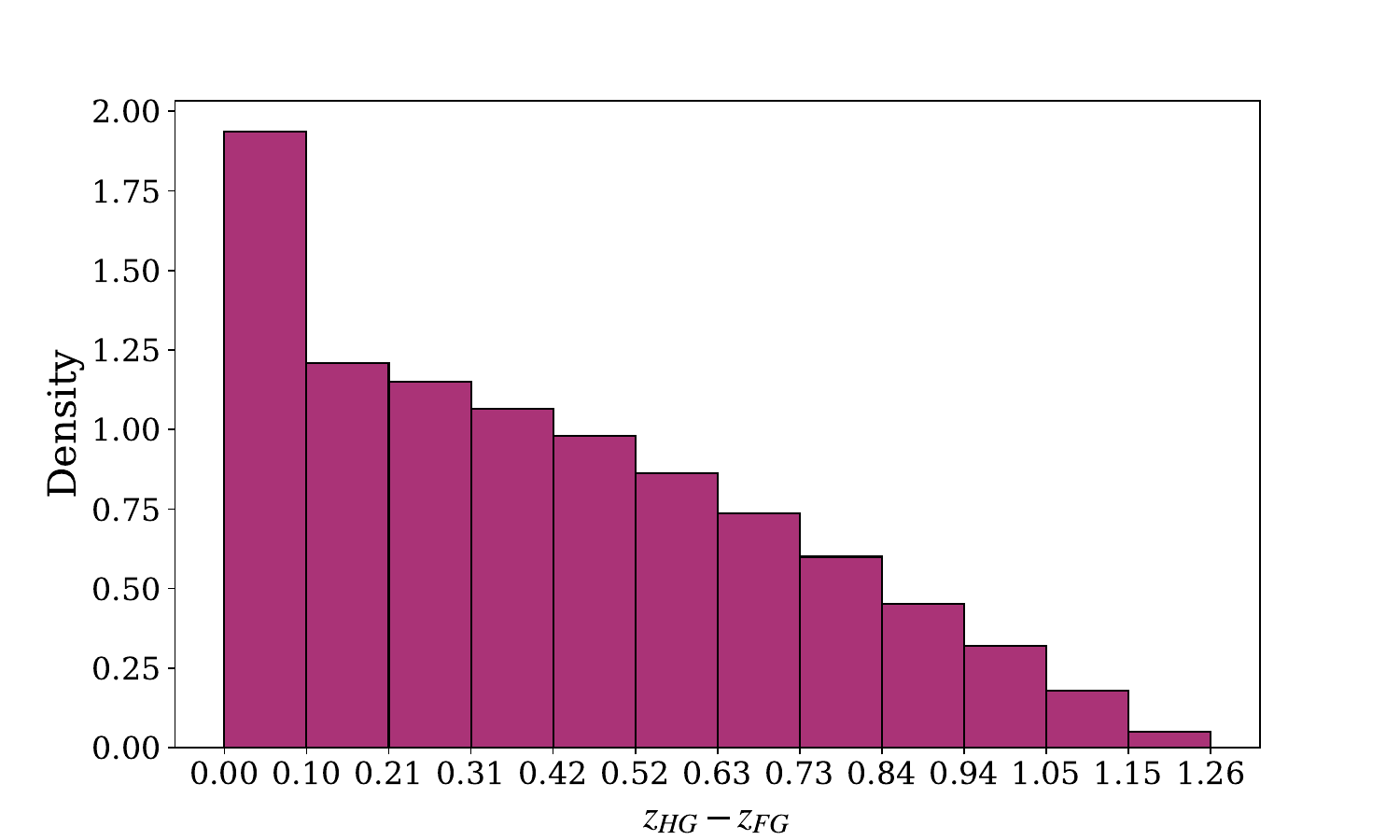}
    \caption{Distribution of $z_{\text{HG}} - z_{\text{FG}}$ for the blocked SN  and foreground galaxy pairs found in our catalogue.}
    \label{fig:z_true_z_obs_dist}
\end{figure}

For mismatched SNe, we examined the difference between the redshift of their true host galaxy and that of their incorrectly matched foreground galaxy, $z_{\text{HG}} - z_{\text{FG}}$. As shown in Figure \ref{fig:z_true_z_obs_dist}, this distribution peaks in the redshift bin $z_{\text{HG}} - z_{\text{FG}} < 0.1$, suggesting that most host galaxy mismatches occur with neighbouring galaxies, likely within the same cluster, leading to relatively small, albeit non-negligible, redshift errors. The tail of this distribution also calls for attention: For $34 \%$ of the mismatched sources, the redshift error caused by mismatch exceeds $z_{\text{HG}} - z_{\text{FG}} = 0.5$. Unless they are removed from the sample, these SNe are mismatched to a distant foreground galaxy and are likely to distort the observed Hubble diagram in a significant way. We assess the impact of host galaxy mismatching and the associated incorrect redshift estimates presented here on $H_0$ in Section \ref{sec:mismatching}. 

\section{\label{sec:method}$H_0$ estimation method}
In this section, we describe the method we used to estimate $H_0$ from a sample of simulated SNe. Late-time estimations of the Hubble constant make use of independent measurements of the luminosity distance and the redshift of SN Ia samples. Under the assumption of the FLRW metric, these two quantities are related through the $H_0$-dependent redshift-distance relation $D(z|H_0)$, which can be displayed in the form of a Hubble diagram. In other words, estimating $H_0$ means fitting $D(z|H_0)$ to the observed Hubble diagram. The FLRW metric does not account for the effects of large-scale structure on the propagation of light, however. These inhomogeneities bias the observed Hubble diagram and must be taken into account when analysing observed SNe Ia samples. We specifically discuss how we addressed the biases introduced in the observed Hubble diagram by gravitational lensing and  peculiar motions to achieve an unbiased estimation of $H_0$.

\subsection{\label{sec:gravitational_lensing}Correction for gravitational lensing}
Gravitational lensing compromises our distance measurements by causing magnified sources (on over-dense lines of sight) to appear closer than they really are and demagnified sources (on under-dense lines of sight) farther away. The apparent distance to sources at fixed redshift therefore is a fluctuating quantity in an inhomogeneous Universe. This translates into a redshift-dependent scatter in the Hubble diagram: At low redshifts, where the line of sight to the observed source is shorter, the dispersion in the Hubble diagram is mainly dominated by peculiar velocities, whereas the effect of lensing becomes increasingly significant for distance measurements at higher redshifts. 

For an observed flux, the probability distribution function (PDF) associated with lensing is skewed and non-Gaussian, with a negative mode and a long positive tail. Photon conservation ensures, however, that the average of this PDF matches the flux received in a homogeneous FLRW Universe \citep{Holz2005}, such that the average flux of a large sample at constant redshift converges to the unlensed flux from which we can derive a true distance measure. This property, in combination with the central limit theorem, can be used to mitigate the scatter introduced by lensing in the Hubble diagram: We binned the observed SNe by redshift and computed the mean redshift and distance for each bin to obtain Gaussianised redshift-distance pairs that form an unbiased Hubble diagram \citep{Adamek2019}.

In addition to binning the SN samples, it is important to use the appropriate distance indicator to construct the unbiased Hubble diagram \citep{Fleury2017}. Since lensing preserves the mean flux density of sources at constant redshift, only distance indicators that are linear functions of the flux density will have the same average value as in a homogeneous Universe. The luminosity distance $d_L$ is related to the flux density $F$ by the cosmological inverse square law $F \propto d_L^{-2}$. Therefore, as put forward in \citet{Adamek2019}, we used the following distance indicator to construct the Hubble diagram,
\vspace{\dimexpr 2\columnsep\relax}
\begin{widetext}
\begin{equation}\label{eq:unbiased_dist_long}
\bar{D} = \bar{d}_L^{-2} = \left[ \frac{1+z}{H_0 \sqrt{\vert\Omega_\mathrm{k}\vert}} f\left(\sqrt{\vert\Omega_\mathrm{k}\vert} \int \limits_0^z \frac{dz'}{\sqrt{\Omega_{\text{r}}(1+z)^4+\Omega_{\text{m}}(1+z')^3+\Omega_{\text{k}}(1+z)^2+\Omega_{\Lambda}}}\right) \right]^{-2}\,,
\end{equation}
\end{widetext}
where $\Omega_{\text{r}}$, $\Omega_{\text{m}}$, $\Omega_{\text{k}}$, and $\Omega_{\Lambda}$ correspond to the density parameters of radiation, matter, curvature, and the cosmological constant, respectively, and $f(x) = \sinh(x), x$, or $\sin(x)$ for an open, flat, or closed spatial geometry. Under the assumptions that $\Omega_{\text{r}} \approx 0$ and $\Omega_{\text{k}} = 0$, which describes a flat geometry, we are left with the following distance indicator that only depends on the cosmological parameters $\theta = \{H_0, \Omega_{\text{m}}\}$:
\begin{equation}\label{eq:unbiased_dist_short}
    \bar{D}(z|\theta) = \Bigg [ \frac{1+z}{H_0} \int \limits_0^z \frac{dz'}{\sqrt{\Omega_{\text{m}}(1+z')^3+(1-\Omega_{\text{m}})}} \Bigg ]^{-2}\,.
\end{equation}
We fitted this relation to the data to determine the biases in the cosmological parameters caused by the selection effects we studied here.

\subsection{\label{sec:peculiar_motions} Mitigation of peculiar motions}
The estimation of $H_0$ from the observed $D(z|H_0)$ relation should rely on the cosmological redshift of SNe Ia, which results solely from the expansion of the Universe. In an inhomogeneous Universe, however, peculiar motions on various scales are associated with peculiar redshifts, which cannot be disentangled from the cosmological redshift in the total redshift that we measure. On large scales, host galaxies that collectively fall into larger clusters have correlated peculiar velocities that result in correlated systematic errors in certain regions of the Hubble diagram. On smaller scales, individual galaxies have random virialized motions relative to their cluster centre of mass associated with uncorrelated peculiar velocities that introduce a random scatter in the Hubble diagram. 
\begin{figure}
    \centering
    \includegraphics[width=0.85\linewidth]{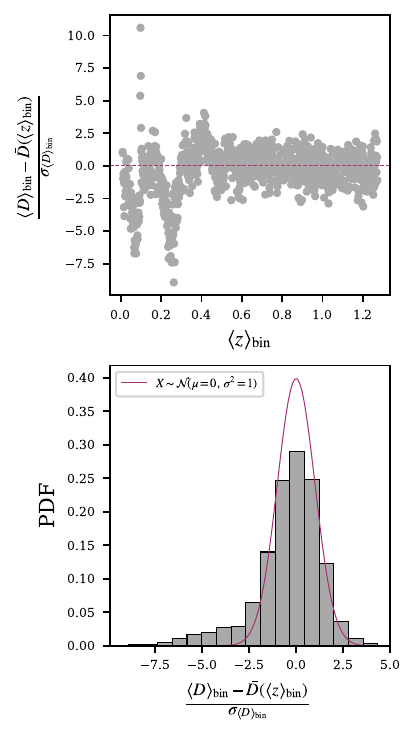}
    \caption{Residuals from the MCMC analysis of a redshift-binned SN sample spanning $0.001<z<1.27$. Top panel: Residuals as a function of redshift. For $z<0.5$, the residuals exhibit a distinct spatial correlation that is attributed to peculiar velocities from large-scale coherent flows. Bottom panel: PDF of the residuals, showing a slight non-Gaussianity due to a negative skew.}
    \label{fig:test_MCMC_residuals}
\end{figure}

Peculiar redshift contamination is particularly significant at low redshift: At $z \sim 0.01$, the radial component of the peculiar velocities of galaxies can account for up to 10\% of the total observed recession velocity \citep{Peterson2022}. The effect of correlated peculiar velocities in coherent flows is also more notable at low redshifts, where the physical separation between sources, as a function of the angular separation on the sky, is smaller. As reported by \citet{Davis2019}, a redshift bias of $5 \times 10^{-4}$ at low redshift can propagate to a bias in $H_0$ of 1 km s$^{-1}$ Mpc$^{-1}$. Moreover, neglecting the correlations between SNe peculiar motions leads to underestimating the uncertainty in the inferred cosmological parameters. Therefore, biases from peculiar motions in the low-redshift part of the Hubble diagram cannot be overlooked and must be corrected for.

Different methods have been developed to correct for the various forms of peculiar motions in our observations \citep{Davis2019}. For the purpose of this study, however, we adopted a simpler approach and mitigated this bias by introducing a low-redshift cutoff in our catalogue of simulated sources. Specifically, we chose to use only the SNe in the redshift range $z > 0.5$ for our estimations of $H_0$. The decision to apply a redshift cutoff at $z = 0.5$ was driven by empirical considerations rather than by a theoretical prescription. We initially performed a Markov chain Monte Carlo (MCMC) analysis (more details are provided in Section \ref{sec:mcmc}) on a redshift-binned SNe Ia sample spanning the full range $0.001 < z < 1.27$ without restrictions, and examined the residuals of the estimated distance measure $D$, defined in Eq. ~\eqref{eq:unbiased_dist_short}, as a function of redshift. Specifically, we considered residuals of the form
\begin{equation}
    \Delta D = \frac{\langle D \rangle_{\text{bin}} - \bar{D}(\langle z \rangle_{\text{bin}})}{\sigma_D},
\end{equation}
where $\langle D \rangle_{\text{bin}}$ is the average estimated distance in a given redshift bin, and $\bar{D}(\langle z \rangle_{\text{bin}})$ is the theoretical value at the average redshift of the bin. The error $\sigma_D$ corresponds to the error on the estimated mean, which for a redshift bin containing $N$ sources is given by
\begin{equation}\label{eq:sigma_D_bin}
    \sigma_D = \frac{\sigma_{\text{bin}}}{\sqrt{N}} = \frac{\sqrt{\sum_i^N (D_i -  \langle D \rangle_{\text{bin}})^2/(N-1)}}{\sqrt{N}}.
\end{equation}
These residuals, shown in Figure \ref{fig:test_MCMC_residuals}, reveal a clear spatially correlated structure for $z<0.5$. These correlations produce a visibly non-Gaussian residual distribution, in contrast to the well-behaved residuals observed at $z>0.5$. We attribute this low-redshift residual structure to correlated peculiar velocities arising from coherent flows. In supernova analyses, these peculiar velocities are typically modelled to mitigate biases in low-redshift measurements. To avoid introducing the complexity of this modelling, we instead imposed a low-redshift cutoff. We acknowledge that this cutoff is stringent and reduces the overlap of our simulated sample with current and forthcoming supernova surveys. Moreover, at redshifts higher than 0.5, the distance–redshift relation $D(z | H_0 )$ becomes increasingly model dependent because the expansion history is dominated by the matter and dark energy density terms. Any inference of $H_0$ in this regime is therefore conditional on the assumed cosmological model, which in this work is a spatially flat $\Lambda$CDM model. We deliberately accepted this model dependence as a trade-off for a reduced sensitivity to peculiar velocity systematics. Our aim was not to replicate any specific observational survey or method, but to isolate and quantify the irreducible systematic bias introduced by foreground galaxies under idealised conditions.

\subsection{\label{sec:mcmc} Markov chain Monte Carlo sampling}
To find the best-fit parameters for $H_0$ and $\Omega_{\text{m}}$ from a set of simulated SNe Ia distance and redshift measurements, we performed an MCMC using the following uncorrelated Gaussian likelihood function:
\begin{equation}\label{eq:likelihood}
        L(D|\theta) = \prod_{\text{SN}} \frac{1}{\sqrt{2\pi}\sigma_D} \, \exp \Bigg[ -\frac{(D - \bar{D})^2}{2 \sigma_D^2} \Bigg]\,.
\end{equation}
In this likelihood function, $D$ corresponds to the distance measured from the simulation, and $\bar{D}$ corresponds to the distance computed according to Eq.~\eqref{eq:unbiased_dist_short} using the measured redshift. The product runs over the average $ \langle D \rangle_{\text{bin}} = \langle d_L^{-2} \rangle_{\text{bin}}$ of each redshift bin in the SN sample, as discussed in Section \ref{sec:gravitational_lensing}.  The error $\sigma_D$, as defined in Eq.~\eqref{eq:sigma_D_bin}, corresponds to the error on this mean. 

For each estimation of $H_0$ that we present in the following sections, we selected a random sample of 300\,000 sources from our simulated catalogue. These sources were then divided into 1000 redshift bins of equal width (with $\Delta z_{\mathrm{bin}} \sim 10^{-4}$). Bins with fewer than 100 sources were discarded from the subsequent analysis as they are not large enough for the central limit theorem to properly Gaussianise their mean. For our simulated catalogue, these correspond at most to a few out of the 1000 bins, depending on the specific randomly selected sources, such that this has no incidence on the subsequent estimation of $H_0$. We opted to bin sources using a constant bin width in redshift rather than a constant source count to ensure that the Hubble diagram was uniformly sampled across the full redshift range we considered, especially at its extremities, where sources are scarcer. This is particularly important because the fit of the Hubble diagram is very sensitive to these extreme points. Therefore, while this decision comes at the cost of losing some information on the redshift distribution of the sample,  it prevents deviations in the inferred $H_0$ due to binning effects. 

The number of sources included in our analysis is significantly larger than that in typical SNe Ia catalogues available today. This deliberate choice allowed us to study the irreducible systematic biases due to SN blocking that persist despite the high quality of the data. Similarly, we assumed vanishing measurement errors on the redshift and luminosity distance of individual sources, such that the scatter in the Hubble diagram is entirely due to the Doppler and gravitational-lensing effects discussed above. 

To perform the MCMC sampling under these conditions, we used the \texttt{Python} package \texttt{emcee} developed by \citet{Emcee2013} to implement the affine-invariant MCMC ensemble sampler developed by \citet{Goodman2010} into \texttt{Python}. We used uniform distributions to define flat priors for $H_0$ and $\Omega_{\text{m}}$: $H_0 \sim U[10,1 00]\,\mathrm{km}\,\mathrm{s}^{-1}\,\mathrm{Mpc}^{-1}$, and $\Omega_{\text{m}} \sim U[0,1]$.

\section{\label{sec:results} Results}

\subsection{\label{sec:total_blocking}Effect of total blocking}

This section is dedicated to determining the effect of the selection bias caused by the total blocking scenario presented in Section \ref{sec:blocking_statistics} on the estimation of $H_0$. About $0.18 \%$ of SNe Ia above a redshift of $0.5$ were excluded from our synthetic catalogue by blocking. We estimated $H_0$ according to the method described in Section \ref{sec:method} using three different SNe Ia samples. The first two were control samples: We selected two different random samples from the simulated catalogue without regard for blocking. They contained both blocked and visible sources. To construct the third sample, we considered that all SNe Ia lying within the half-light radius of a foreground galaxy are not visible to us, and we only selected sources that were visible according to this criterion. The sample accounting for SN blocking led to a slightly higher central value for $H_0$ and a slightly lower central value for $\Omega_{\text{m}}$ than the two control samples. These biases are not significant at a statistical level, however, which is expected because the fraction of blocked SNe Ia is small.

A small caveat regarding the low-redshift cutoff introduced in Section \ref{sec:peculiar_motions}: As the fraction of blocked SNe increases with redshift, this cutoff might slightly amplify the effect of SN blocking on $H_0$ by focusing on the part of the Hubble diagram that is affected most by SN blocking. On the other hand, the lower mass threshold we imposed to construct the halo catalogue from the N-body simulation caused us to underestimate the density of small galaxies when we performed the abundance matching, and consequently, also the fraction of blocked sources. Therefore, taking these two effects into account, it is reasonable to assume that the effect of SN blocking on $H_0$ presented here remains conservative and is likely to be somewhat more pronounced when using observed SN catalogues. The systematic bias is still expected to remain irrelevant because of the precision of current $H_0$ estimates. This selection bias might require further investigation in the coming years, however, if we significantly increase the size and redshift range of SN samples.

\subsection{\label{sec:mismatching}Effect of incorrect host assignments}
In this section, we assess the impact of incorrect host galaxy assignments and the corresponding redshift error presented in Section \ref{sec:mismatching_statistics}. We estimate $H_0$ twice using the same SN sample and following the method described in Section \ref{sec:method}. For the first estimation, we assigned to each SN the redshift of its true host galaxy, assuming that no obstructed SN was mismatched to a foreground galaxy. For the second estimation, we assigned the redshift of the foreground galaxy to all SNe that were matched to a foreground galaxy during the implementation of the matching algorithm developed by \citet{Gupta2016}. These represent $\sim 2.9\%$ of the sources in our simulated catalogue at $z > 0.5$.

\begin{figure*}
    \sidecaption
    \includegraphics[width=0.65\linewidth]{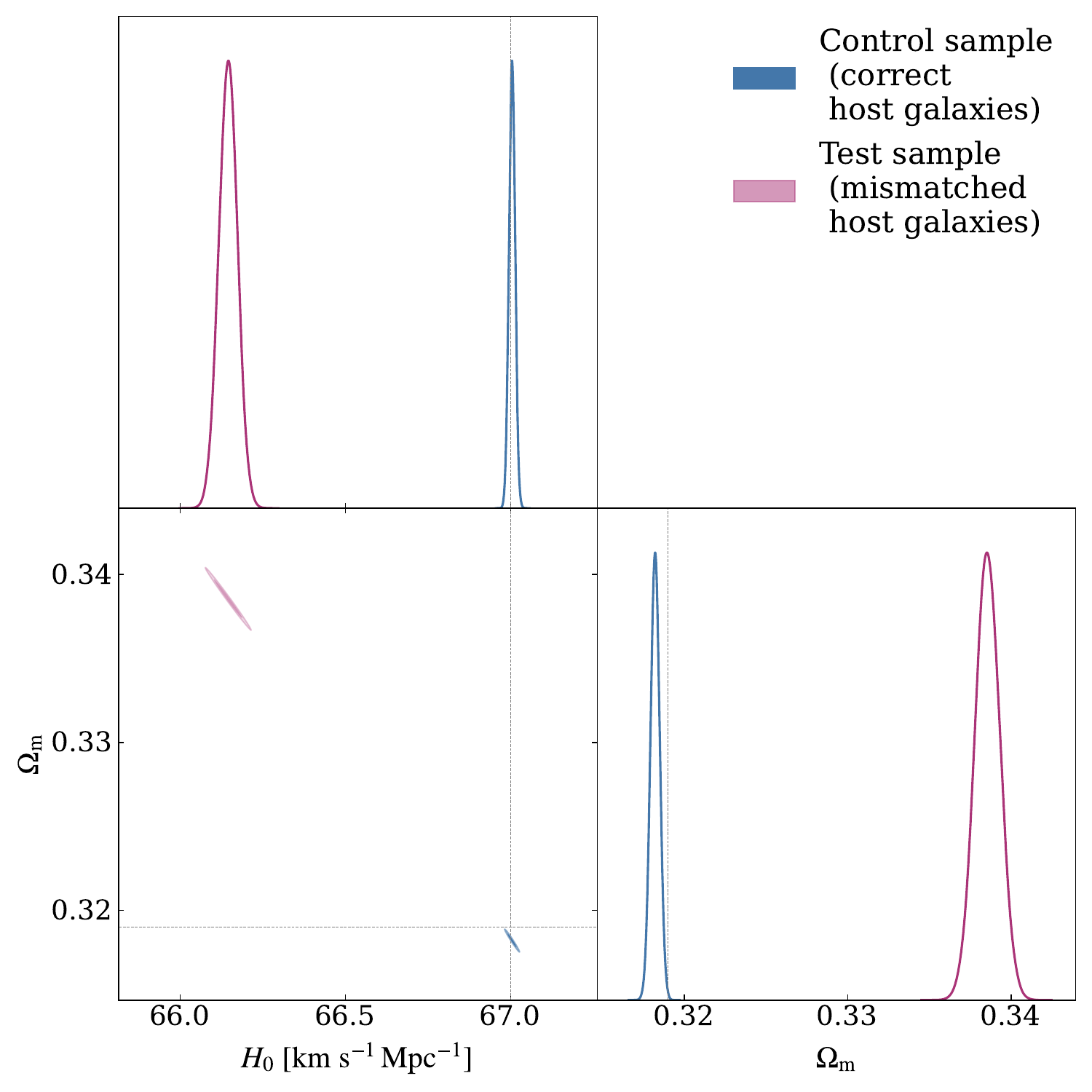}
    \caption{Constraints on $H_0$ and $\Omega_{\text{m}}$ for two different scenarios of the host galaxy assignment. The blue contours correspond to the control case, where all SNe Ia are correctly matched to their host galaxies. The purple contours represent the sample in which some SNe Ia are assigned to an incorrect host galaxy because the matching criterion relies solely on angular separation. The dashed grey line indicates the true parameter values we used in the simulation.}
    \label{fig:mismatching_MCMC_cornerplots}
\end{figure*}

The estimates for $H_0$ and $\Omega_{\text{m}}$ resulting from these two estimations are shown in Figure \ref{fig:mismatching_MCMC_cornerplots}. The comparison of these estimates reveals that the incorrect host galaxy assignment of obstructed SNe Ia causes us to underestimate $H_0$ and overestimate $\Omega_{\text{m}}$. More specifically, the $H_0$ estimate from the sample with mismatched SNe is $ \sim 1.3 \%$ lower than that obtained from the control case (which corresponds in absolute terms to a difference of $\sim 0.86 \,\mathrm{km}\,\mathrm{s}^{-1}\,\mathrm{Mpc}^{-1}$). In their latest measurement, the \textit{SH0ES} team reached a sub-percent level on each of the error components, which combined to a total error of  $1.35 \%$ \citep{Riess2021}. 
This total error is of the same order of magnitude as the shift in $H_0$ found here to be caused by SN host galaxy misidentification. The \textit{SH0ES} analysis mainly concerns SNe much below $z = 0.5$ where mismatching is less likely (see Figure~\ref{fig:mismatching_vs_z}), and they can benefit from the fact that many of their SNe have direct spectroscopic redshifts to check for this systematic. It is apparently already relevant to account for this effect, however, based on the precision of current $H_0$ estimates: \citet{Carr2022} visually inspected all low-redshift and some moderate-redshift SNe in the sample and reported six examples with ambiguous host assignment, which is roughly in line with our expectation. Moving forward, this systematic is bound to gain significance in the coming years as our measurement accuracy improves and our observations extend to higher redshifts.

We note that the $2\sigma$ contour of the posterior distribution obtained from the control sample does not encompass the true values of the cosmological parameters we used in the simulation, indicating that there is a slight residual bias in the constructed binned and averaged Hubble diagram. Although \citet{Adamek2019} achieved an unbiased estimation of cosmological parameters using a similar binning approach, their analysis included SNe up to $z=3$ and bins of 1000 sources. Therefore, we suspect that increasing the bin sizes might improve the accuracy of the $H_0$ estimation by enhancing the Gaussianising effect of the central limit theorem, and including sources at higher redshift would further reduce the bias from peculiar motions. It would be interesting to explore further how this binning procedure might be optimised to achieve an unbiased $H_0$ estimation while using the smallest possible SNe Ia sample. 

\begin{figure}
    \centering
    \includegraphics[width=1\linewidth]{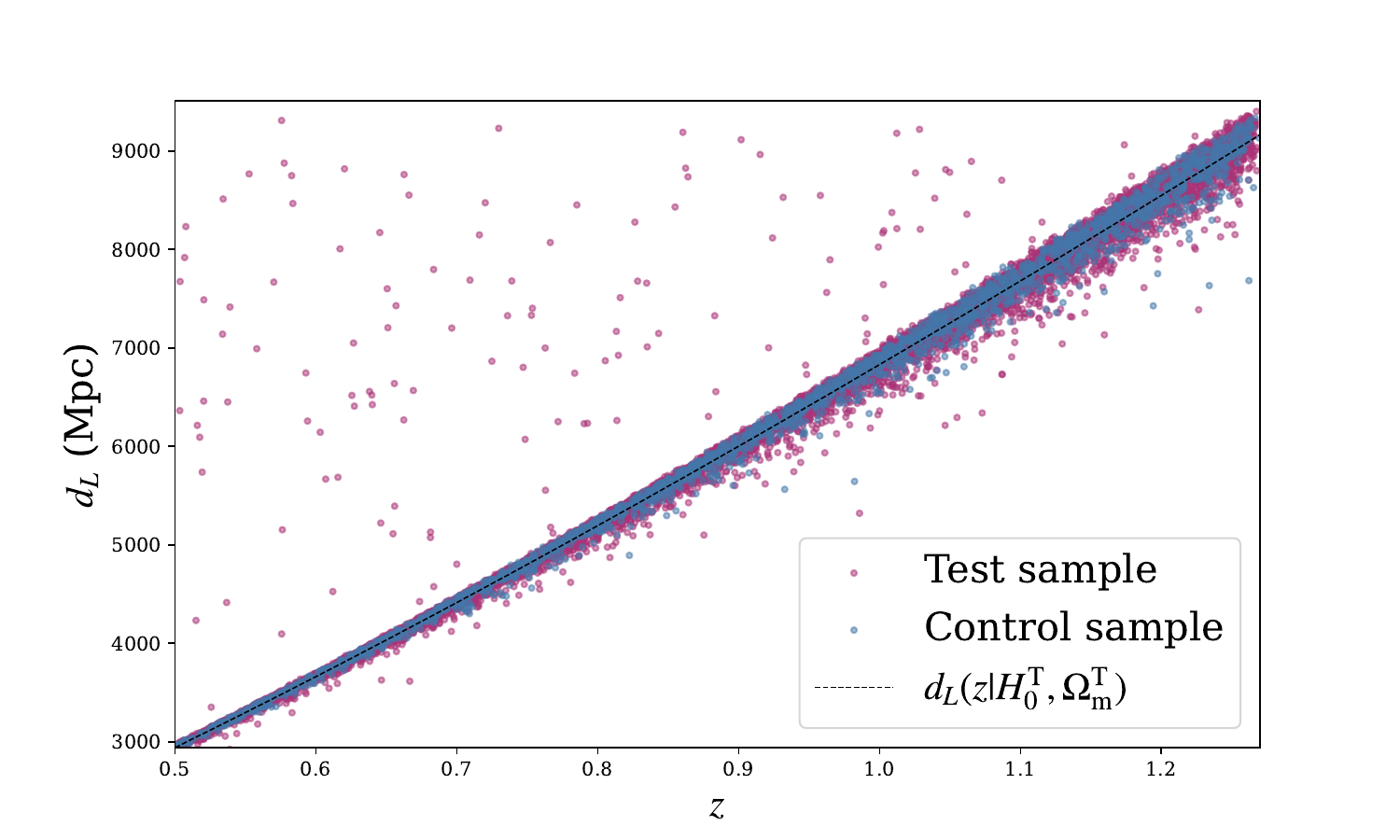}
    \caption{Hubble diagram constructed from two different SNe samples: In the case shown in blue, all SNe are correctly matched to their respective host galaxies. In contrast, in the case shown in purple, the host galaxies were matched to each SNe Ia with the algorithm described in Section \ref{sec:mismatching_statistics}. The dashed $d_L(z)$ curve was computed with the parameters fitted from the sample where all SNe Ia are correctly matched to their host galaxies. We see by comparison that the SNe for which the host galaxy has been misidentified are scattered across the upper left region of the plot, shifted horizontally to lower $z$ along a line of constant $d_L$.}
    \label{fig:Mismatching_HubbleDiagram}
\end{figure}

The decrease in the estimated value of $H_0$ can be understood by analysing the impact of the host galaxy misidentifications in the Hubble diagram. Figure \ref{fig:Mismatching_HubbleDiagram} shows the Hubble diagram from the SN sample we used in the second inference, where the matching algorithm described in Section \ref{sec:mismatching_statistics} was applied. The SNe for which the host galaxy was misidentified are scattered across the upper left region of the plot, shifted horizontally to lower $z$ along a line of constant $d_L$. These sources therefore introduce a significant bias in the Hubble diagram towards higher values for $d_L$ at fixed $z$, particularly at low redshifts. These mismatched SNe pull the fitted $d_L$ upwards at fixed $z$, yielding a lower $H_0$ and higher $\Omega_{\text{m}}$ value. The Hubble diagram shown in Figure \ref{fig:Mismatching_HubbleDiagram} was constructed using the SNe Ia samples from which we inferred the $H_0$ and $\Omega_{\text{m}}$ constraints in Figure \ref{fig:mismatching_MCMC_cornerplots}, each consisting of $300\,000$ sources. This large sample size highlights the effect of host galaxy mismatching and explains why it is so noticeable although it affects only about $2.9\%$ of SNe. For typical observed SNe Ia samples, such as the \textit{Pantheon}$+$ catalogue, which includes $O(10^3)$ sources \citep{Scolnic2022}, we expect approximately $O(10)$ sources to be mismatched. In this case, incorrect host galaxy assignments might not be easily visible in the observed Hubble diagram.
 
As described in Section \ref{sec:mismatching_statistics}, the host galaxy matching algorithm we used relies on proximity in projected distance between the SNe Ia and its host galaxy, while excluding any potential hosts with an angular separation greater than $30''$ from the SN from the matching process. \citet{Gupta2016} reported a $97 \%$ accuracy for this algorithm based on tests performed using a simulated catalogue of $\sim 100\,000$ SNe Ia placed onto two different samples of observed galaxies in the redshift range $0.08 < z < 1.4$. They also observed that the fraction of mismatched SNe increased with redshift, which means that the accuracy of the algorithm depends on redshift \citep[see][top left panel of Fig.\ 9 and left panel of Fig.\ 10]{Gupta2016}. Furthermore, most mismatches occurred with galaxies at lower redshifts than the true hosts, leading to a skewed $z_{\text{true}} - z_{\text{match}}$ distribution \citep[see][right panel of Fig.\ 10]{Gupta2016}.

These findings raise several important considerations, which are particularly relevant in the context of this study on the misidentification of obstructed SNe host galaxies. First, all SNe located at more than $30''$ from their host galaxies were removed from the simulated sample by \citet{Gupta2016}. Although this represents a small subset of sources ($0.05 \%$ in their first galaxy catalogue and $0.6 \%$ in the second), their exclusion likely led to a slight overestimation of the algorithm accuracy. Due to their large separation from their host centres, these SNe are more likely to be mismatched to a foreground galaxy when they are obstructed. The $2.6 \%$ of obstructed SNe Ia that were mismatched to foreground galaxies, as identified in our analysis, might at least partially account for the $3 \%$ inaccuracy reported by the authors in their matching algorithm. This interpretation is supported by the redshift dependence of the algorithm accuracy and the skewed distribution of $z_{\text{true}} - z_{\text{match}}$, as the number of obstructed SNe increases with redshift and foreground galaxies have lower redshifts than the true hosts. Although \citet{Gupta2016} did not provide an estimate of the effect of host galaxy misidentification on the inferred cosmological parameters, the results presented here show that the mismatch of obstructed SNe Ia alone can already shift $H_0$ quite significantly. Considering these observations alongside our results, the host galaxy matching algorithm might be improved, and consequently, the accuracy of $H_0$ estimates, by factoring in SN obstruction by foreground galaxies.

\section{\label{sec:conclusion}Conclusion}
With the forthcoming launch of surveys such as the \textit{Legacy Survey of Space and Time (LSST)} by the \textit{Vera C. Rubin Observatory} and the \textit{James Webb Space Telescope (JWST)}, we anticipate a significant increase in the size of observed SNe Ia samples and their redshift span. This will likely shift the dominant source of error in $H_0$ estimates from statistical to systematic errors and emphasises the need for a more robust consideration of smaller systematic biases so that we can continue to improve our late-time constraints on $H_0$. This work assessed how foreground galaxies that obstruct the lines of sight to certain supernovae in our observations can bias our estimation of the Hubble constant.

Using a synthetic SNe Ia catalogue, we showed that fewer than $1 \%$ of high-redshift supernovae are potentially obstructed by a foreground galaxy. These few blocked sources are, however, more magnified than their visible counterparts, however, because they tend to lie along over-dense lines of sight. In its simplest form, this blocking excludes certain supernovae from observed samples. This selection bias leads to a slight overestimation of $H_0$ that is not statistically significant given the uncertainties in the latest constraints on $H_0$. 

To date, efforts to improve the SNe Ia distance ladder have primarily focused on enhancing the accuracy of supernova luminosity measurements, and thus, their use as reliable standard candles. In contrast, errors arising from the indirect determination of supernova redshifts via their host galaxies have received less attention and were often regarded as negligible. By applying the host galaxy matching algorithm used in the latest \textit{SH0ES} $H_0$ analysis to our synthetic catalogue, we found that $2.6\%$ of SNe Ia are incorrectly matched to a foreground galaxy. Most of these mismatches involve a neighbouring foreground galaxy within the same cluster, although a significant fraction of sources ($34\%$) are mismatched to a galaxy with a redshift difference of $z_{\text{HG}} - z_{\text{FG}} > 0.5$. These incorrect host galaxy assignments result in an underestimated $H_0$ by $1.3\%$ in our simulation. The latest $H_0$ measurement by the \textit{SH0ES} team is unlikely to be affected to this extent because the majority of supernovae in their sample have independent spectroscopic redshift measurements and lie at $z<0.5$, where the probability of host mismatching is very low. We stress, however, the importance of further investigating the impact of host galaxy misidentification in current observations and future higher-redshift surveys and of developing matching algorithms that can mitigate these biases.

The overall bias on $H_0$ resulting from line-of-sight obstruction is a combination of two effects: total blocking, and incorrect host galaxy assignment, the latter being the dominant effect in our study. We therefore conclude that further efforts to improve the accuracy of our late-time estimation of the Hubble constant should incorporate the consideration of host misidentification of obstructed supernovae in the matching algorithm and should correct for the selection bias induced by total blocking when larger observed SNe Ia samples become available.

\begin{acknowledgements}
      The work of JA is supported by the Swiss National Science Foundation. This work was supported by a grant from the Swiss National Supercomputing Centre (CSCS) under project ID s1035.
\end{acknowledgements}

\bibliographystyle{aa}
\bibliography{aa54139-25}

\end{document}